\documentclass[11pt, a4paper]{article}
\pdfoutput=1
\usepackage[T1]{fontenc}
\usepackage[utf8]{inputenc}
\usepackage[english]{babel}
\usepackage{amsmath}
\usepackage{amssymb}
\usepackage{graphicx}
\usepackage{xcolor, color}
\usepackage{multirow}

\usepackage{url}
\usepackage{xspace}
\usepackage{subcaption}
\usepackage{jcappub}
\usepackage{natbib}
\usepackage{comment}
\usepackage{bm}
\usepackage{braket}
\usepackage{amssymb}
\usepackage{amsmath}

\newcommand{\fnl}{f_{\rm NL}}

\definecolor{light-gray}{gray}{0.70}

\newcommand*\diff{\mathop{}\!\mathrm{d}}
\newcommand*\Diff[1]{\mathop{}\!\mathrm{d^#1}}
\renewcommand{\vec}{\bm}
\newcommand{\threej}[6]{ \begin{pmatrix}
    #1 & #2 & #3 \\
    #4 & #5 & #6
\end{pmatrix}}
\newcommand{\sixj}[6]{ \begin{Bmatrix}
    #1 & #2 & #3 \\
    #4 & #5 & #6
\end{Bmatrix}}

\graphicspath{{./Immagini/}}

\title{Primordial non-Gaussianity with $\mu$-type and $y$-type spectral distortions: exploiting Cosmic Microwave Background polarization and dealing with secondary sources}

\author[a,b,1]{Andrea Ravenni,\note{Corresponding author.}}
\author[a,b,c]{Michele Liguori,}
\author[a,b,c]{Nicola Bartolo}
\author[d,e]{and Maresuke Shiraishi}

\affiliation[a]{Dipartimento di Fisica e Astronomia “G. Galilei”, Università degli Studi di Padova, via~Marzolo~8, I-35131, Padova, Italy}
\affiliation[b]{INFN, Sezione di Padova, via~Marzolo~8, I-35131, Padova, Italy}
\affiliation[c]{INAF-Osservatorio Astronomico di Padova, vicolo~dell'Osservatorio~5, I-35122 Padova, Italy}
\affiliation[d]{Department of General Education, National Institute of Technology, Kagawa College,\\355~Chokushi-cho, Takamatsu, Kagawa 761-8058, Japan}
\affiliation[e]{Kavli Institute for the Physics and Mathematics of the Universe (Kavli IPMU, WPI), UTIAS, The University of Tokyo, Chiba, 277-8583, Japan}

\emailAdd{ravenni@pd.infn.it}
\emailAdd{liguori@pd.infn.it}
\emailAdd{bartolo@pd.infn.it}
\emailAdd{shiraishi-m@t.kagawa-nct.ac.jp}

\abstract{Cross-correlations between Cosmic Microwave Background (CMB) temperature and $y$-spectral distortions anisotropies have been previously proposed as a way to measure the local bispectrum parameter $\fnl^{loc.}$ in a range of scales inaccessible to either CMB ($T$, $E$) bispectra or $T$-$\mu$ correlations. This is useful e.g. to test scale dependence of primordial non-Gaussianity. Unfortunately, the primordial $y$-T signal is strongly contaminated by the late-time correlation between the Integrated Sachs Wolfe and Sunyaev-Zel'dovich (SZ) effects. Moreover, SZ itself generates a large noise contribution in the $y$-parameter map. We consider two original ways to address these issues. In order to remove the bias due to the SZ-CMB temperature coupling, while also adding new signal, we include in the analysis the cross-correlation between $y$-distortions and CMB {\em polarization}.
In order to reduce the noise, we propose to clean the $y$-map by subtracting a SZ template, reconstructed via cross-correlation with external tracers (CMB and galaxy-lensing signals). We combine this SZ template subtraction with the previously adopted solution of directly masking detected clusters. Our final forecasts show that, using $y$-distortions, a PRISM-like survey can achieve $1\sigma(\fnl^\text{loc.}) = 300$, while an ideal experiment will achieve $1\sigma(\fnl^\text{loc.}) = 130$, with improvements of a factor $\sim 3$ from adding the $y$-$E$ signal, and a further $20-30 \%$ from template cleaning. These forecasts are much worse than current $\fnl^\text{loc.}$ boundaries from {\em Planck}, but we stress again that they refer to completely different scales.}

\begin{document}
\maketitle
\flushbottom

\section{Introduction}

At present, the tightest constraints on all parametrizations and models of primordial non-Gaussianity (NG) come from measurements of the bispectrum (Fourier transform of the 3-point function in configuration space) of Cosmic Microwave Background (CMB) temperature and polarization anisotropies (respectively $T$ and $E$), made by the \textit{Planck} satellite \cite{Ade:2015ava}.

Among many aspects and applications of these constraints, a very important one is the possibility to set stringent bounds on inflationary scenarios characterized by more than one field. Multi-field Inflation in fact predicts a potentially detectable bispectrum of the local type, peaking in the so-called squeezed-limit (i.e., on wavenumber triangles with one side much smaller than the other two, indicating a correlation between large and small wavelengths). Such bispectrum explicitly reads: 
\begin{equation}
B(k_1, k_2, k_3)=-\frac{6}{5}f_\text{NL}^{\rm loc.}\left[P(k_1)P(k_2)+\text{2 perm.}\right] ,
\label{eq:LocalBispectrum}
\end{equation}
where $f_\text{NL}^{\rm loc.}$ is the dimensionless local NG amplitude parameter, which is measured by fitting the local shape to the data (since we will consider only local NG 
in the following, we will omit the superscript ``loc.'' from now on). Currently, \textit{Planck} constrains $f_\text{NL} =0.8 \pm 5.0$ (68\% C.L.) \cite{Ade:2015ava}.  A crucial threshold 
to fully distinguish single from multi-field scenarios would be however $\fnl \sim 1$. This value is in fact a lower bound for a large class of multi-field models (e.g. curvaton \cite{Bartolo:2003jx}). Unfortunately there are not enough modes in the CMB temperature and polarization angular bispectra to achieve enough sensitivity for a clear detection of $\fnl \sim 1$, 
even assuming a perfectly noiseless, ideal survey (see, e.g. \cite{Finelli:2016cyd}).
Several new observational strategies will therefore have to be implemented in the future.
One of the most promising approaches in the near future involves measuring NG signatures in the galaxy bias, using forthcoming Large Scale Structure surveys \cite{Dalal:2007cu, Matarrese:2008nc}.

In a more futuristic scenario, the authors of \cite{Pajer:2012vz} have considered the cross-correlation between CMB temperature and $\mu$-spectral distortion {\em anisotropies} as a potentially very powerful probe of squeezed-type bispectra.
An experiment producing cosmic-variance limited $T$ and $\mu$ maps could in principle be able to detect $\fnl \sim 10^{-2}, \, 10^{-3}$ (this argument has been extended in different ways \cite{Ganc:2012ae, Biagetti:2013sr, Miyamoto:2013oua, Kunze:2013uja, Ganc:2014wia, Ota:2014iva,  Shiraishi:2015lma, Emami:2015xqa, Bartolo:2015fqz, Ota:2016mqd, Creque-Sarbinowski:2016wue}, to take into account different primordial bispectrum models and higher-order correlation functions).  While fascinating, this scenario is out of reach not only with current experimental noise levels, but also taking into account proposed, ambitious next generation surveys, 
such as PRISM \cite{Emami:2015xqa}.

Nonetheless, measurements of $\fnl$ via correlations between CMB temperature (polarization) and CMB distortion anisotropies are interesting even before achieving such exquisite levels of sensitivity, as they allow to test possible deviations 
of $\fnl$ from scale independence. In this respect, not only $\mu$, but also $y$-distortions can provide useful and interesting information, as pointed out in \cite{Emami:2015xqa}.
Indeed, while \textit{Planck} measured $f_\text{NL}$ on the scales typical of CMB $T$, $E$ anisotropies ($k\lesssim 0.15$ Mpc$^{-1}$), and $\mu$-distortion anisotropies can in principle probe it on much smaller scales ($46$ Mpc$^{-1} \lesssim k \lesssim 10^4 $ Mpc$^{-1}$), $y$-distortions allow probing the transition between those two regimes ($0.15$ Mpc$^{-1} \lesssim k \lesssim 46$ Mpc$^{-1}$).
The issue with using $y$-distortions is, however, that the total signal is affected by contributions from secondary sources, which completely dominate over the primordial component. The most important source of contamination is the $y$-$T$ signal generated 
by the correlation between the Integrated Sachs Wolfe (ISW) and Sunyaev-Zeld'ovich (SZ) effects.
This problem has been initially addressed for the $y$-$T$ cross-correlation in \cite{Creque-Sarbinowski:2016wue}, where a cosmic-variance limited experiment was considered, and the SZ contamination was reduced by masking detected 
clusters at low redshift.
In this paper, we will extend previous analyses by considering not only $y$-$T$, but also including the cross-correlation with polarization, $y$-$E$, and by exploiting 
cross-correlations between SZ and external tracers (CMB and galaxy lensing).
Besides that of adding new signal, using polarization presents the clear advantage of giving a much less biased signal, since the $E$-mode correlate less than $T$ with the SZ effect.
Nevertheless, we will have in this case to worry about potential spurious contamination from reionization. An explicit numerical evaluation, using second-order transfer functions from the Boltzmann integrator \texttt{SONG} \cite{pettinari:2013a,pettinari:2014a,pettinari:2015a} will show that this is negligible.\footnote{\url{https://github.com/coccoinomane/song.}}
The correlation between SZ and lensing can instead be used to estimate a template of the $y$-parameter map generated by SZ, which can then be subtracted from the data, in order to partially remove spurious SZ contributions from 
unresolved clusters and reduce the noise. We will consider a PIXIE-like \cite{Kogut:2011xw}, PRISM-like \cite{Andre:2013nfa} and an ideal, cosmic-variance limited experiment, and show how including these new ingredients can lead to interesting improvements in the final forecasts, by an overall factor $\sim 4$ in all cases.

While we will focus mostly on $y$-distortions, we will also extend previous $\mu$-distortions/ E-polarization cross-correlation analyses. More specifically, we will re-analyse in detail the $\mu$-$E$ cross-spectrum
initially discussed in \cite{Ota:2016mqd}, where only large scales and reionization contributions to the CMB polarization transfer functions were included. In that case, it was found that  $\mu$-$E$ does not provide any further 
constraining power with respect to $\mu$-$T$. By considering all scales and using full transfer functions we will obtain that $\mu$-$E$ performs slightly better than $\mu$-$T$ and combining the two leads to $\sim 20 \%$ 
improvement in the final constraints.

Even though we are interested in a $f_\text{NL}=f_\text{NL}(k_1,k_2,k_3)$, which depends on scale, 
we will follow \cite{Emami:2015xqa, Creque-Sarbinowski:2016wue} to assume that the scale-dependence is such that $\fnl$ stays approximately 
constant (separately) on both the $y$-scales ($0.15$ Mpc$^{-1} \lesssim k \lesssim 46 $ Mpc$^{-1}$) and the $\mu$-scales ($46$ Mpc$^{-1} \lesssim k \lesssim 10^4 $ Mpc$^{-1}$).

The paper is organized as it follows:
In section \ref{sec:PrimordialSignal} we will calculate the primordial contributions to the cross correlation of $T$ and $E$ with $\mu$- and $y$-spectral distortions.
in section \ref{sec:Foregrounds} we will calculate the secondary sources for the $T$ and $E$ cross correlations with $y$ --- we recall that $\mu$ does not have cosmological secondary sources.
In section \ref{sec:MuForecast} and \ref{sec:YForecast} we will forecast $f_\text{NL}^\mu$ and $f_\text{NL}^y$  constraints, achievable by a PIXIE-like, a PRISM-like, and by a cosmic-variance limited survey, considering all sources of noise and contamination, different masks for resolved clusters (based on future X-ray and CMB surveys) and different external tracers 
for unresolved contributions. In section \ref{sec:Conclusions} we will summarize our conclusions.

\section{Primordial contributions}
\label{sec:PrimordialSignal}
In this section we review the calculation to obtain the cross-correlation of CMB temperature and polarization anisotropies with $\mu$- or $y_p$-CMB-spectral-distortions anisotropies, when primordial NG is present.
Since secondary sources generate $y$-distortion in the late universe we indicate with $y_p$ the primordial contribution to the total $y$.

The primordial primordial curvature perturbation $\zeta (\vec{k})$ and CMB fluctuations are linked via
\begin{equation}
a_{\ell m}^X=4\pi i^\ell \int \frac{\Diff{3} \vec{k}}{(2\pi)^3}\mathcal{T}_\ell^X (k) Y_{\ell}^{m*} (\vec{\hat{k}}) \zeta (\vec{k}),
\end{equation}
where $X=T,E$ indicate the temperature $T$ or the $E$-mode polarization, and $\mathcal{T}_\ell^X$ is the radiation transfer function.
We will use the full transfer function generated by \texttt{CLASS} \cite{Lesgourgues:2011re}.
For  $\Xi = \mu$-$, y_p$-type spectral distortions the analogous relation reads instead  \cite{Ganc:2012ae, Bartolo:2015fqz}
\begin{equation}
\begin{split}
a^\Xi_{\ell m} = 4\pi (-i)^\ell \int& \frac{\Diff{3} \vec{k_1}}{(2\pi)^3}\frac{\Diff{3} \vec{k_2}}{(2\pi)^3} \Diff{3}\vec{k_3} \delta^{(3)}(\vec{k_1}+\vec{k_2}+\vec{k_3})
\\
& Y_\ell^{m*}(\vec{\hat{k}_3}) j_\ell(k_3 r_\text{ls}) f^\Xi(k_1,k_2,k_3)\zeta (\vec{k_1})\zeta (\vec{k_2}),
\\
f^\mu(k_1,k_2,k_3) =& 2.3 \ W\!\left(\frac{k_3}{k_D(z_{\mu y})}\right)\left[e^{-(k_1^2+k_2^2)/k_D^2(z)}\right]^{z_\mu}_{z_{\mu y}}
\\
f^y(k_1,k_2,k_3) =& 0.4 \ W\!\left(\frac{k_3}{k_D(z_y)}\right)\left[e^{-(k_1^2+k_2^2)/k_D^2(z)}\right]^{z_{\mu y}}_{z_{y}}.
\end{split}
\end{equation}
Here $r_\text{ls}$ is the comoving distance to last scattering surface, $W(x)=3j_1(x)/x$, and $k_D$ is the diffusion damping scale evaluated at the beginning of the $\mu$-era $k_D(z_\mu)\approx \, 12000\text{ Mpc}^{-1}$, at the $\mu$-$y$ transition $k_D(z_{\mu y})\approx \, 46\text{ Mpc}^{-1}$, and at the end of the $y$-era $k_D(z_y)\approx \, 0.15\text{ Mpc}^{-1}$ \cite{Emami:2015xqa}.
As said, we consider $f_\text{NL}$ constant on $\mu$- and $y$-scales: $f_\text{NL}(k) \equiv f_\text{NL}^y$ for $0.15$~Mpc$^{-1}\lesssim k \lesssim 46$~Mpc$^{-1}$ and $f_\text{NL}(k) \equiv f_\text{NL}^\mu$ for $46$~Mpc$^{-1}\lesssim k \lesssim 12000$~Mpc$^{-1}$.

More accurate expressions for the transfer function have been dicussed in \cite{Chluba:2016aln}.
We will use the simpler approximations, but we will re-normalize the amplitude of the expected monopoles of $\mu$- and $y$-distortion to the values computed in \cite{Chluba:2016bvg},
$\langle \mu \rangle = 2.3 \times 10^{-8}$ and $\langle y \rangle = 4.2 \times 10 ^{-9}$
.
This ``zero-order'' approximation is accurate enough for a Fisher forecast and allow a simple, direct comparison with other results in the literature.

The cross-correlation can be found as:

\begin{equation}
\begin{split}
\Braket{a_{\ell m}^X a^{\Xi*}_{\ell' m'}}= & 8\ \delta_\ell^{\ell'}\delta_m^{m'}\int \frac{k^2\diff k}{2\pi^2}\mathcal{T}_\ell^X (k)  j_{\ell'}(k r_\text{ls})
\int \frac{q_1^2 q_2^2\diff q_1\diff q_2}{2\pi^2} 
 f^\Xi(q_1,q_2,k)
B(k,q_1, q_2)
\\
& \int x^2 \diff x j_0(q_1 x)j_0(q_2 x)j_0(k x).
\end{split}
\label{ClXi-X_exact}
\end{equation}

In \cite{Shiraishi:2015lma} it has been discussed how 
the generation mechanism for $\mu$ via acoustic dissipation,
encoded in the $f^\mu(q_1,q_2,k)$ function,
strongly selects squeezed configuration $k_1\approx k_2\gg k$ in the $T$-$\mu$ correlation.
The same argument holds for the $E$-$\mu$ cross-correlation.
Since the diffusion damping scale of $y_p$ is much smaller than the $\mu$ one, the same approximation is less accurate in the $X$-$y_p$ cross-correlations.
However it should be noted that the more accurate transfer function provided in \cite{Chluba:2016aln} explicitly suppress configuration with too different $q_1$ and $q_2$.
This means that also when using $y$-distortions we are allowed to take the squeezed limit $k_1\approx k_2\gg k$.

In the squeezed limit, eq. (\ref{ClXi-X_exact}) reduces to
\begin{equation}
C_\ell^{X\Xi} \approx -4\pi \frac{12}{5} \int \frac{k^2\diff k}{2\pi^2}\mathcal{T}_\ell^X (k)  j_{\ell'}(k r_\text{ls})
P(k) \int \frac{q_1^2 \diff q_1}{2\pi^2} 
 f^\Xi(q_1,q_1,k)
P(q_1).
\label{ClXi-X_squeezed}
\end{equation}
Notice that the last integral in this equation is exactly the definition of the monopole of the $\Xi$-type spectral distortion. 
Thus, renormalizing the $\langle \mu \rangle$ and $\langle y \rangle$ to the right vale as discussed above translate linearly into a renormalization of the $C_\ell$

\begin{figure}
\centering
\includegraphics[width=0.7\textwidth]{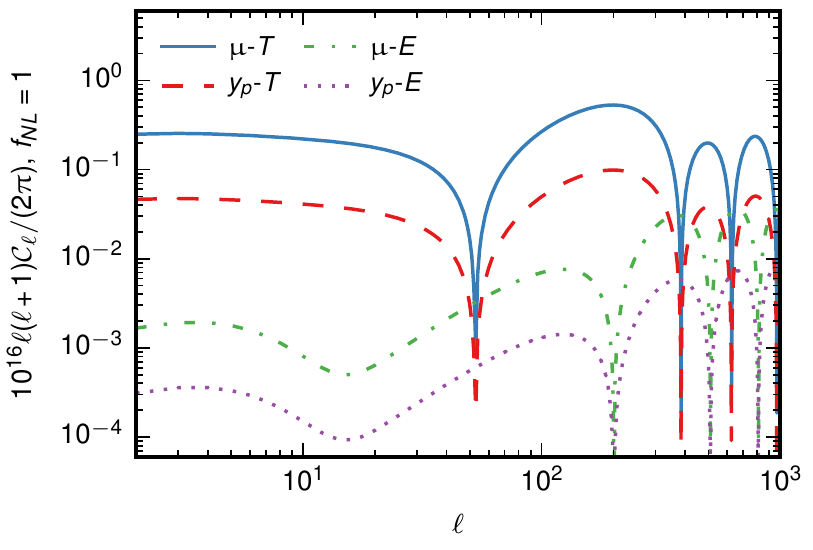}
\caption{Absolute value of the cross-correlations. The prediction for $\mu$-$T$ is in good agreement with \cite{Shiraishi:2015lma}.}
\label{fig:Signal}
\end{figure}

In figure \ref{fig:Signal} we show the cross-correlations between $\mu$- or $y_p$-distortion, and $T$ or $E$ anisotropies.
The prediction for $\mu$-$T$ is in good agreement with \cite{Shiraishi:2015lma}.
Only the primordial contribution to the cross-correlations is shown in the plot,
even though other effects contribute to the same signals.
We are going to consider them in the next section, since those secondary sources will constitute foregrounds to these primordial signal.

\section{Secondary sources}
\label{sec:Foregrounds}

In this section we will consider the main non-primordial contributions to the $y$-$T$ and $y$-$E$ spectra, namely the cross-correlation between Sunyaev-Zel'dovich (SZ) and the Integrated-Sachs-Wolfe (ISW) effect, for $y$-$T$, and the cross correlation between CMB polarization and the quadratic Doppler effect, for $y$-$E$.

\subsection{Sunyaev-Zel'dovich}

The SZ effect generates $y$-distortions that couple to CMB temperature anisotropies produced via late ISW.
This has been studied at length in \cite{PhysRevD.65.103510, 2011MNRAS.418.2207T, Creque-Sarbinowski:2016wue}, and references therein.

To calculate the SZ we use a halo model approach, following  \cite{2011MNRAS.418.2207T, Hill:2013baa, Komatsu:2002wc}.
We parametrize the density of dark matter haloes in term of the matter overdensity distribution $\delta$, using a bias parameter $b(z,M)$, which depends on redshift and mass of the halo.
The mass distribution of haloes is given in terms of the halo mass function $\frac{\diff n}{\diff M}(z,M)$.
Since the SZ is sensitive to the electron rather than to the matter distribution, this has to be convolved with the halo Compton $y$-parameter image $y_{3D}(z,M,x)$, where $x$ is the distance from the center of the halo; $y_{3D}$ is a function of the electron pressure profile of the halo.
We consider respectively the bias given in table 2 of \cite{2010ApJ...724..878T}, the halo mass function of \cite{Tinker:2008ff} with the updated parameters given in \cite{2010ApJ...724..878T}, and the halo Compton $y$-parameter computed in \cite{Battaglia:2011cq}.

The one- and two-halo terms \cite{Hill:2013baa,2011MNRAS.418.2207T} respectively read
\begin{gather}
C_\ell^{1h}  = \int \diff z \frac{\Diff{2} V}{\diff z \diff \Omega} \int \diff M \frac{\diff n}{\diff M} (z, M) |\tilde{y}_\ell(z,M)|^2
\label{eq:Clyy1h}
\\
C_\ell^{2h}  = \int \diff z \frac{\Diff{2} V}{\diff z \diff \Omega} D_+^2(z) P_m(k) \bigg[\int \diff M \frac{\diff n}{\diff M} (z, M)  b(z, M) \tilde{y}_\ell(z,M) \bigg]^2\bigg|_{k=\big(\frac{\ell + 1/2}{\chi(z)}\big)}\; .
\label{eq:Clyy2h}
\end{gather}
Here $P_m(k)$ is the linear matter power spectrum, $D_+(z)$ is the growth factor, $\Diff{2} V/ \diff z \diff \Omega = c\chi^2(z)/H(z) $ is the comoving volume element per steradians and $\tilde{y}_\ell(z,M)$ is the 2D Fourier transform of the projected $y$-parameter image of the halo
\begin{equation}
\tilde{y}_\ell(z,M) = \frac{4\pi r_{s,y}}{\ell_{s,y}^2} \int \diff x x^2 j_0\left(\frac{k x}{\ell_s}\right) y_{3D}(z,M,x),
\end{equation}
$r_{s,y}$ is the typical scale radius of the $y$-image of the halo and $\ell_{s,y} = a(z) \chi(z)/r_{s,y} $.

We refer to the appendix of \cite{Hill:2013baa} for a clear derivation of these two formulae.
For an explicit numerical evaluation of these integrals, we worked in Limber approximation. This allows removing one of the $5$ nested 
integrals, making the computation  numerically feasible.

The SZ effect cross correlate with $T$ through the late ISW effect, given by
\begin{equation}
\frac{\Delta T^\text{ISW}}{T}(\vec{\hat{n}})=-\frac{2}{c^2}\int \diff z \frac{\diff\phi}{\diff z} (\chi(z) \vec{\hat{n}},z).
\end{equation}
In our numerical evaluation, we will use the full transfer functions, extracted from \texttt{CLASS} \cite{Lesgourgues:2011re}, which of course encode this contribution.

\noindent It is useful to define real space transfer functions \cite{Komatsu:2003fd, Liguori:2003mb}
\begin{equation}
\mathcal{T}^T_\ell (\chi) = \frac{2}{\pi}\int \diff k \, k^2 \mathcal{T}^T_\ell (k) j_\ell(\chi k) \; ,
\end{equation}
where $\mathcal{T}^T_\ell (k)$ is the full temperature transfer function.
The multipolar coefficients can then be expressed as
\begin{equation}
a_{\ell m}^T = \int \diff \chi\,  \chi^2 \mathcal{T}^T_\ell (\chi) \zeta_{\ell m} (\chi) \; ,
\end{equation}
where $\zeta_{\ell m} (\chi) = \frac{(-i)^\ell}{2\pi^2} \int \Diff{3}\vec{k} \,\zeta (\vec{k}) j_\ell(\chi k) Y_\ell^m(\vec{\hat{k}})$.

Using the Poisson equation, $\phi (\chi(z) \vec{\hat{n}},z)= -\frac{3}{2}\Omega_{M}H_0^2 \frac{D(z)}{a(z)} \mathcal{T}_m (k) \nabla^{-2} \delta(\chi(z) \vec{\hat{n}}, z=0)$, to express the overdensity contrast as a function of the gravitational potential we find
\begin{equation}
\begin{split}
C_\ell^{\text{SZ-}T} =&\int \frac{c \diff z}{H(z)} \frac{3}{5}\frac{c^2 k^2 \chi^2(z) \mathcal{T}_m (k)}{\Omega_m H_0^2}D_+(z)\int \diff M \frac{\diff n}{\diff M}(z,M) \tilde{y}_\ell\big(z,M) b(z,M)\times
\\
&
\times\mathcal{T}^T_\ell (\chi(z))
P_\zeta (k)\bigg|_{k=\big(\frac{\ell + 1/2}{\chi(z)}\big)}.
\end{split}
\label{eq:SZT}
\end{equation}

As previously mentioned, one of the main goals of this paper is to study the $\fnl$ dependence in the $y$-$E$ cross-correlation spectrum, which was not accounted for in previous works on the subject.
One of the main advantages of using $y$-$E$ in place of, or in combination with, $y$-$T$, is that contamination from SZ is expected 
to be strongly suppressed for $y$-$E$, therefore the main source of bias that afflicts $y$-$T$-based measurements of $\fnl$ \cite{Creque-Sarbinowski:2016wue} would be eliminated.
While much smaller than the $y$-$T$ contribution, a non-primordial $y$-$E$ correlation is still present: after reionization the quadrupole of free electrons still acts as a source of $E$. SZ is generated in the same epoch on similar scales, so this gives rise to a non-vanishing $y$-$E$.
This is expected to be a very small effect at low redshifts. However, also the primordial $y_p$-$E$ signal we are after is very small,
therefore, it is important to explicitly compare the two effects.

The $C_\ell^{\text{SZ}E}$ can be computed replacing the transfer functions for $T$ with those for $E$ in eq. (\ref{eq:SZT}). Even though, practically, all the SZ signal comes from $z<4$ \cite{Hill:2013baa}, since $E$ is sourced at reionization we extend all the redshift integrations to $z$ well above the time of reionization. As expected, the contributions from $z>4$ are negligible.
As a lower integration limit we choose $z>0.02$; this ensure that the redshift integrals do not get contributions from unphysical $z = 0$ objects.
We also integrate over the masses $10^{10}M_\odot h^{-1} < M < 10^{16}M_\odot h^{-1}$.
We checked our spectra against those shown in \cite{Hill:2013baa},
changing our integration boundaries to match their choices, and we are in very good agreement.

\subsection{$y$-distortion from reionization}

Another source of contamination comes from the cross-correlation of $E$ with the quadratic Doppler effect (see \cite{Renaux-Petel:2013zwa, Hu:1995em} and references therein).
Being proportional to the velocity of the baryons squared, this observable tracks the primordial density squared $y_\text{reio} \sim \zeta(k)^2$.
It is then clear that its cross-correlation with first order CMB polarization anisotropies is proportional to the primordial bispectrum $C_\ell^{y_\text{reio}\text{-}E} \propto \Braket{\zeta (\vec{k_1})\zeta (\vec{k_2})\zeta (\vec{k_3})}$.
However, we neglect this potential contribution to the signal here, leaving its study for future work, and focus instead on spurious, non-primordial contamination, which need to be removed from the primordial $y$-$E$ contribution at recombination.

To simplify the notation we will omit ``$C_\ell$'' in defining cross-correlations (i.e., $y_\text{reio}\text{-}E \equiv C_\ell^{y_\text{reio}\text{-}E}$) and we will indicate the $n$-th order term of a quantity with an apex ``$(n)$'', i.e., $E^{(n)}$.

Assuming perfectly Gaussian primordial perturbations, the leading term of the quadratic Doppler effect-Polarization cross-correlation is of fourth order in the primordial density perturbation: $y_\text{reio}\text{-}E = y_\text{reio}^{(2)}\text{-}E^{(2)} + y_\text{reio}^{(3)}\text{-}E^{(1)} + \mathcal{O}(\zeta^6)$.
We expect the second addendum to be of the same order of the first.
Calculating it would require developing new formalism to describe higher-order-contributions to spectral distortions.
Since, as we anticipate, we found after a complete calculation that the first term is negligible, we neglect the calculation of the second.

The authors of \cite{Renaux-Petel:2013zwa} provide an analytic expression for $y_\text{reio}^{(2)}$. The second order transfer function for CMB polarization anisotropies can instead be obtained numerically, using the publicly available code \texttt{SONG} \cite{pettinari:2013a,pettinari:2014a,pettinari:2015a}.
We found that the cross-correlation between these two quantities is (see appendix \ref{sec:yE2fullcalculation} for the full calculation)

\begin{equation}
\begin{split}
y_\text{reio}^{(2)}\text{-}E^{(2)} = 
&(-1)^{\ell'-m}64\pi
\int \frac{q_1^2 \diff q_1 q_2^2 \diff q_2}{(2\pi)^3}
\int \frac{k_1^2\diff k_1}{(2\pi)^3}
P(q_1)P(q_2)\delta_{\ell}^{\ell'}\delta_{m}^{-m'}
\\
&
\bigg[
\overline{\mathcal{T}}_{X\ \ell, 0}^{(2)}(q_1, q_2, k_1)
\frac{1}{3}I_{\ell'}^{(1)}(q_1,q_2, k_1)
\int x^2 \diff x 
j_{0}(x k_1)j_{1}(x q_1)j_{1}(x q_2) +
\\
&
+\sum_{L\ L_1}\sum_{m_1}\sum^1_{n=-1} 
\overline{\mathcal{T}}_{X\ \ell m_1}^{(2)}(q_1, q_2, k_1)
\frac{11\pi}{45}I_{\ell',m_1}^{(2)}(q_1,q_2, k_1)
(-1)^{L_1+1}
i^{L+L_1+1}
(-1)^{3m_1}
\\
&
\threej{L_1}{1}{L}{0}{0}{0}
\threej{L_1}{1}{L}{n}{-n}{0}
\threej{L_1}{|m_1|}{1}{0}{0}{0}
\threej{L_1}{|m_1|}{1}{-n}{m_1}{-m_1+n}
\\
&
\frac{3(2L+1)(2L_1+1)}{4\pi}
\alpha_{n,m_1}
\int x^2 \diff x 
j_{L}(x k_1)j_{L_1}(x q_1)j_{1}(x q_2)
\bigg].
\end{split}
\end{equation}
To calculate this signal we use the transfer function extracted from \texttt{SONG} with 10\% accuracy.
Since the convergence of the tensor modes has not been tested by the authors for more than 10\% accuracy \cite{pettinari:2015a}, using higher precision run would require extensive testing of the code. Moreover, as the final $y^{(2)}$-$E^{({2})}$ contribution will turn out to be negligible, our results will not depend on this quantity, making accuracy improvements not important for our purposes.

\begin{figure}
\centering
\includegraphics[width=0.7\textwidth]{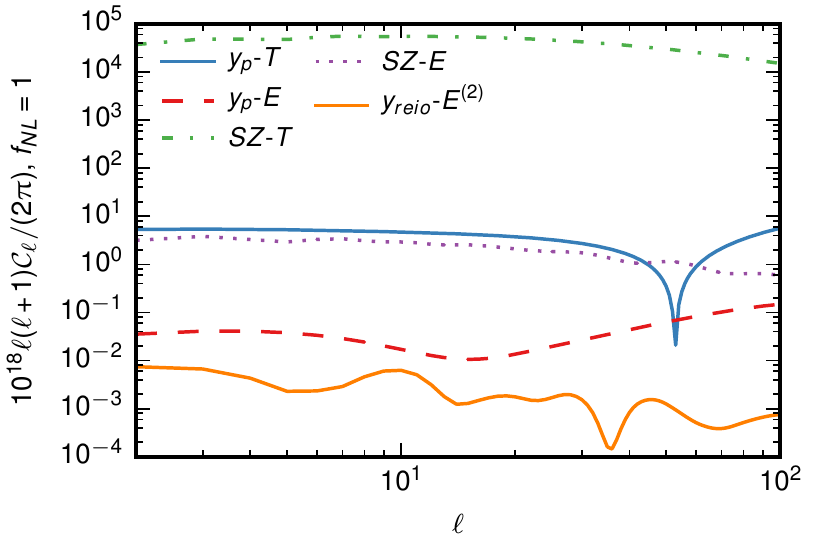}
\caption{The main components (primordial and secondary) of the $y$-$T$ and $y$-$E$ cross-correlations. $SZ$-$T$ is the only non-negligible contaminant to $y_p$-$T$, but is four orders of magnitude bigger, and correlates more with it. $y_p$-$E$ suffers the presence of both $SZ$-$E$ and $y_\text{reio}$-$E$, but both secondary signals are proportionally smaller and less correlated than $SZ$-$T$.}
\label{fig:yreioE2}
\end{figure}

\medskip

In figure \ref{fig:yreioE2} we compare the $y_p$-$E$ signal with the secondary sources of $E$-$y$.
The SZ-$E$ cross-correlation is approximately 100 times bigger than the signal, while the $y_\text{reio}$-$E^{(2)}$ cross-correlation is 10-100 times smaller than the signal in the first 100 multipoles, and their ratio decreases as $\ell$ increases.
The slightly different slopes allow disentangling of the signal from the secondary sources, as we will show in section \ref{sec:YForecast}.
For comparison, the SZ-$T$ cross-correlation in $10^4$ bigger than the primordial signal.
The same marginalization over the foregrounds can be performed also in this case, but with worse results since the shape of the primordial signal is more degenerate with the shape of the secondary sources.
This can be quantified calculating the correlation of the primordial signal with the secondary signal:
\begin{equation}
\frac{\Big[\sum_\ell C_\ell^{y_p T} C_\ell^{\text{SZ} T}\Big]}{\sqrt{\Big[\sum_\ell \big(C_\ell^{y_p T} \big)^2\Big] \Big[\sum_\ell \big(C_\ell^{\text{SZ} T}\big)^2\Big]}}=0.92,
\quad 
\frac{\Big[\sum_\ell C_\ell^{y_p E} C_\ell^{\text{SZ} E}\Big]}{\sqrt{\Big[\sum_\ell \big(C_\ell^{y_p E} \big)^2\Big] \Big[\sum_\ell \big(C_\ell^{\text{SZ} E}\big)^2\Big]}}=0.64.
\label{eq:SZT-yTcorr}
\end{equation}

\section{$f_\text{NL}^\mu$ forecast}\label{sec:forecastmu}
\label{sec:MuForecast}

As first considered in \cite{Ota:2016mqd}, we can try enhancing the $\fnl$ signal-to-noise ratio, extracted using $T$-${\mu}$, by adding polarization to the analysis.
In \cite{Ota:2016mqd}, simplified ``Sachs-Wolfe''-limit transfer functions were used both for temperature and polarization anisotropies, finding no $\fnl$-sensitivity improvements with the inclusion of polarization. However, we found via 
explicit computation that the inclusion of full transfer functions does change this picture at $\ell > 10$, outside the limits of validity of the Sachs-Wolfe approximation.

The Fisher matrix is defined as 
\begin{equation}
F_{ij}\equiv \Braket{ \frac{\partial L}{\partial{p_i}}\frac{\partial L}{ \partial{p_j}}} \; ,
\label{eq:Fdef}
\end{equation}
where $L$ is the logarithm of the likelihood and ${\bf p}$ are the free parameters of the theory.
In our application it is equivalent to \cite{Heavens:2009nx}
\begin{equation}
F_{ij}= \sum_\ell (\textbf{Cov}^{-1}_\ell)_{\alpha\beta} \frac{\partial (\textbf{Cov}_\ell)_{ \beta\gamma}}{\partial{p_i}}
(\textbf{Cov}_\ell^{-1})_{\gamma\delta} \frac{\partial (\textbf{Cov}_\ell)_{\delta\alpha}}{\partial{p_j}}
\end{equation}
where $\textbf{Cov}_{\ell}$ is the covariance matrix and repeated matrix indices ($\alpha, \dots ,\delta$) are summed.

The  $f^\mu_\text{NL}$ Fisher matrix, when considering only $T$ or only $E$, reads \cite{Pajer:2012vz}
\begin{equation}
\bigg(\frac{S}{N}\bigg)^2=
F=\sum_{\ell=2}^{\ell_\text{max}} \frac{C_\ell^{\mu X}C_\ell^{\mu X}}{(\sigma^{\mu X}_\ell)^2},
\end{equation}
where we recall that $X=T,E$.
For a PIXIE- or PRISM-like experiment it is expected that $C_\ell^{X\mu}C_\ell^{X\mu} \ll (C_\ell^{XX})_\text{obs}(C_\ell^{\mu\mu})_\text{obs},$ and $(C_\ell^{\mu\mu})_\text{obs} \approx (C_\ell^{\mu\mu})_N \gg C_\ell^{\mu\mu}$ where ``obs'' stands for observed and ``$N$'' for noise.
Again, for PIXIE (PRISM) the expected noise is $C_\ell^{\mu\mu,N} = 4\pi \times (1.4\times 10^{-8})^2 \times e^{\ell^2/84^2}$ ($C_\ell^{\mu\mu,N} = 4\pi \times 10^{-18} \times e^{\ell^2/100^2}$) \cite{Ganc:2012ae, Emami:2015xqa}.
Here we do not account for galactic foregrounds.
In \cite{Abitbol:2017vwa} it has been shown that the error on the measurement of the spectral distortion monopole can degrade as much as one order of magnitude with respect to earlier, more optimistic estimates.
The spatially varying part of these foregrounds will have to be modelled with more accuracy in order to give an actual estimate of how much the estimates will degrade in our case.

\begin{figure}
\centering
\includegraphics[width=0.7\textwidth]{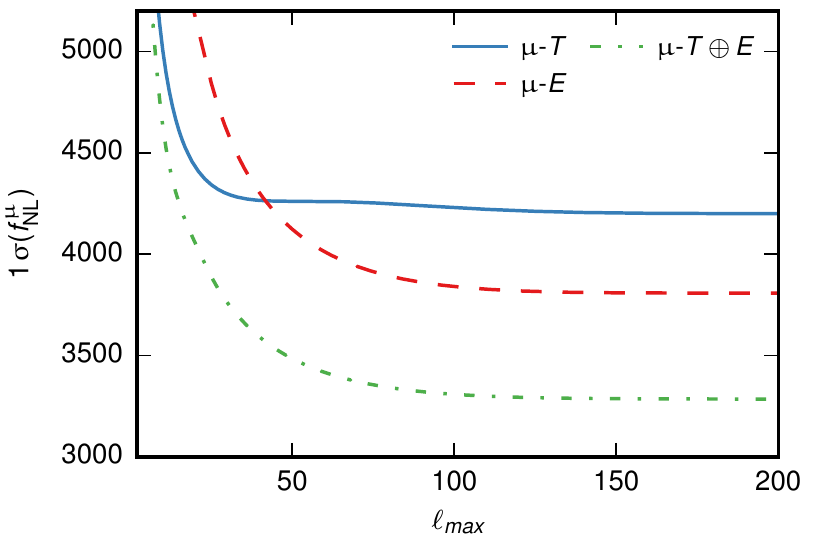}
\caption{Minimum value of $f_\text{NL}$ to reject $f_\text{NL}=0$ at $1\sigma$, as a function of the maximum multipole. This value is calculated with a 1 parameter Fisher forecast using the PIXIE level of noise. Adding polarization informations tightens the constrain by a factor 1.28.
}
\label{fig:MuForecast}
\end{figure}

Under those assumptions
\begin{equation}
(\sigma_\ell^{\mu X})^2\approx \frac{C_\ell^{XX}C_\ell^{\mu\mu,N}}{2\ell + 1}\;.
\label{eq:MuNoise}
\end{equation}
Therefore, the signal-to-noise ratio is proportional to the $\mu E$ contribution --- which is underestimated for $\ell>10$ using the ``Sachs-Wolfe''-transfer function --- and inversely proportional to the square root of the polarization power spectrum, which peaks on the first 10 multipoles (due to the reionization bump) and decreases afterwards.
As a result an explicit numerical evaluation shows that the $\fnl^\mu$ signal-to-noise ratio from $E$-$\mu$ is actually higher than the one obtained using $T$-$\mu$, at $\ell_{\rm max} > 50$, see figure \ref{fig:MuForecast}.

We are finally interested in the joint estimate of $f_\text{NL}^\mu$, obtained combining both observations:
\begin{equation}
F=\sum_\ell^{\ell_\text{max}} (2\ell+ 1) \frac{C_\ell^{TT}(C_\ell^{\mu E})^2 + C_\ell^{EE}(C_\ell^{\mu T})^2 - 2 C_\ell^{TE}C_\ell^{\mu T}C_\ell^{\mu E} }{C_\ell^{\mu\mu,N}[C_\ell^{TT}C_\ell^{EE}-(C_\ell^{TE})^2]}
\label{eq:MuJointForecast}
\end{equation}
as the covariance matrix is
\begin{equation}
\textbf{Cov}_{\ell}
=
\frac{1}{2\ell+1}\begin{pmatrix}
C_\ell^{TT}                           & C_\ell^{TE}                            & f^\mu_\text{NL} C_\ell^{\mu T}\\
C_\ell^{TE}                           & C_\ell^{EE}                            & f^\mu_\text{NL} C_\ell^{\mu E}\\ 
f^\mu_\text{NL} C_\ell^{\mu T}  & f^\mu_\text{NL} C_\ell^{\mu E}  & C_\ell^{\mu \mu, N}\\   
\end{pmatrix}. 
\end{equation}

\begin{table}
\centering
\begin{tabular}{|ccccc|}
\hline
        & Survey    & $T$      & $E$  & $T\oplus E$\\
\hline
\multirow{2}{*}{$1\sigma(f_\text{NL}^\mu)$}
    & PIXIE             & 4200     & 3800 & 3300  \\
    & PRISM             & 300      & 270  & 230  \\
\hline
\end{tabular}
\caption{1$\sigma$ forecasted error bars on $f_\text{NL}^\mu$,
calculated using the standard $\Lambda$CDM value of $\langle \mu \rangle = 2.3 \times 10^{-8}$.
$T\oplus E$ indicates the joint forecast using both temperature and polarization.
We accounted for correlations between $T$-$\mu$ and $E$-$\mu$ using eq. (\ref{eq:MuJointForecast}).}
\label{tab:Muresults}
\end{table}

As expected for a PIXIE-like survey, the signal-to-noise ratio saturates for $\ell \approx 100$. We found that adding the polarization cross-correlation to the temperature cross-correlation with the $\mu$-spectral-distortion the constraint on $f_\text{NL}$ improves by a factor 1.28.
In figure \ref{fig:MuForecast} we show the minimum value of $f^\mu_\text{NL}$ that guarantees a $1\sigma$ rejection of $f^\mu_\text{NL}=0$, as a function of the maximum multipole.
Our results are shown in table \ref{tab:Muresults}.

\section{$f_\text{NL}^y$ forecast}
\label{sec:YForecast}

We will now come to the main point of this work, namely studying the effects of adding $y$-distortions in the $\fnl$ analysis, including contributions from polarization and exploring methods to clean SZ-contamination via SZ-lensing correlations.

\subsection{$T$-y forecast and cluster masking.}

In our forecasts we have to keep into account theoretical uncertainties of the halo-model, used to predict the correlations.
Following \cite{Creque-Sarbinowski:2016wue}, we will do this by simply introducing an unknown amplitude parameter $\alpha_T$, in front of
the spectra, and marginalizing over it.
This leads to the covariance matrix
\begin{equation}
\textbf{Cov}_{\ell}
=
\frac{1}{2\ell+1}\begin{pmatrix}
C_\ell^{TT}                                                                                 & f^y_\text{NL} C_\ell^{y T} + \alpha_T C_\ell^{\text{SZ} T} \\
f^y_\text{NL} C_\ell^{y T} + \alpha_T C_\ell^{\text{SZ} T }   & C_\ell^{yy,N} + C_\ell^{1h} + C_\ell^{2h}\\   
\end{pmatrix}. 
\label{eq:Tmarg}
\end{equation}

A PIXIE-like experiment is expected to have 5 to 10 better sensitivity to $y$ than to $\mu$.
Therefore the noise term is $C_\ell^{yy,N} = 4\pi \times 4 \times 10^{-18} \times e^{\ell^2/84^2}$;
the same holds for PRISM for which $C_\ell^{yy,N} = 4\pi \times 4 \times 10^{-20} \times e^{\ell^2/84^2}$ \cite{Ganc:2012ae, Emami:2015xqa}.
We forecast $f_\text{NL}^y$ using temperature alone, and marginalizing over the secondary source SZ-$T$. 

\begin{figure}
\centering
\includegraphics[width=0.7\textwidth]{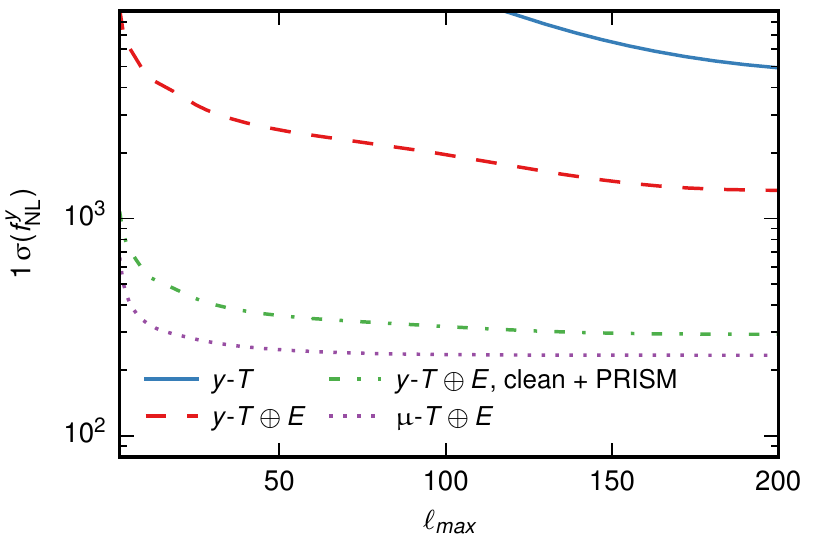}
\caption{Minimum value of $f^y_\text{NL}$ needed to reject $f^y_\text{NL}=0$ at $1\sigma$, as a function of the maximum multipole. This value is calculated with a Fisher forecast after marginalizing over all foregrounds and using the PRISM level of noise.
We also show  the same value for $f^\mu_\text{NL}$ for comparison.
We compare the results obtained using no mask and no template subtraction ($y$-$T \oplus E$) and using the PRISM mask and the template subtraction ($y$-$T \oplus E$, clean + PRISM) described respectively in table \ref{tab:masks} and section \ref{sec:cleaning}.
}
\label{fig:YForecast}
\end{figure}

In figure \ref{fig:YForecast} we show our results as a function of $\ell_\text{max}$. It is clear not only that the variance is completely dominated by the SZ power spectrum, but also that this effect leads to very poor constraints on the primordial signal. 

One way to tighten the constraints is to mask resolved clusters (see e.g \citep{Creque-Sarbinowski:2016wue, Hill:2013baa}), eventually assuming the use 
of external surveys (e.g. X-ray surveys) to improve performance. 
We will consider here eROSITA \cite{Merloni:2012uf} as external survey, and also PRISM itself, and investigate different types of masks, based on more or less futuristic scenarios, in order to understand which level of masking guarantees a signal-to-noise ratio for $f_\text{NL}^y$ similar to the one achieved for $f_\text{NL}^\mu$.
Our results are summarized in table \ref{tab:results}.

\begin{table}
\centering
\begin{tabular}{|cccccc|}
\hline
\multirow{6}{*}{$1\sigma(f_\text{NL}^y)$}
    & Mask                & $T$      & $E$  & $T\oplus E$ &$T\oplus E$, clean.\\
\hline
    & Unmasked            & 12700    & 5500 & 3300        & 2900 \\
    & eROSITA             & 8600     & 4800 & 2700        & 2300 \\
    & PRISM               & 5500     & 4000 & 2200        & 2200 \\
    & $z>0.3$             & 5500     & 4200 & 2300        & 2300 \\
\hline
\end{tabular}
\caption{1$\sigma$ forecasted error bars on $f_\text{NL}^y$ for PIXIE,
calculated using the standard $\Lambda$CDM value of $\langle y \rangle = 4.2 \times 10 ^{-9}$.
$T\oplus E$ indicates the joint forecast using both temperature and polarization.
We accounted for correlations between $T$-$y$ and $E$-$y$ using the covariance in eq. (\ref{eq:YJointForecast}).
In all the forecasts we marginalize over the amplitude of every secondary source.
$z>0.3$ performs worse than PRISM mask because the contribution to the total SZ signal coming from small ($M<10^{13}\ M_\odot$) clusters at low ($z<0.3$) redshift is smaller than that of bigger clusters at higher redshift.}
\label{tab:results}
\end{table}

\begin{table}
\centering
\begin{tabular}{|cccccc|}
\hline
\multirow{6}{*}{$1\sigma(f_\text{NL}^y)$}
    & Mask                & $T$      & $E$  & $T\oplus E$ &$T\oplus E$, clean.\\
\hline
    & Unmasked            &	4900	&	3100	&	1700	&	1300	\\
    & eROSITA             &	3200	&	1900	&	1100	&	680	\\
    & PRISM               &	1000	&	630	    &	380	    &	300	\\
    & $z>0.3$             &	1700	&	1300	&	700	    &	620	\\
\hline
\end{tabular}
\caption{Same as table \ref{tab:results} but for PRISM.}
\label{tab:resultsPRISM}
\end{table}

\begin{table}
\centering
\begin{tabular}{|cccccc|}
\hline
\multirow{6}{*}{$1\sigma(f_\text{NL}^y)$}
    & Mask                & $T$      & $E$  & $T\oplus E$ &$T\oplus E$, clean.\\
\hline
    & Unmasked            &	2300	&	1400	&	1000	&	750	\\
    & eROSITA             &	1700	&	1100	&	730	    &	470	\\
    & PRISM               &	400	    &	220	    &	160	    &	130	\\
    & $z>0.3$             &	1000	&	710	    &	470	    &	400	\\
\hline
\end{tabular}
\caption{Same as table \ref{tab:results} but for a cosmic-variance limited spectrometer instead of PIXIE, and using $\ell_\text{max}=1000$.
}
\label{tab:resultsCV}
\end{table}

We model the effect of masking clusters by changing integration boudaries in the SZ-SZ, SZ-T and SZ-E spectra, in order to exclude regions in the $z$-$M$ plane  where the catalogue of a given experiment is complete \cite{Hill:2013baa}.
This is a very conservative choice, as in real catalogues a non-negligible part of resolved cluster actually sits in regions where the catalogue is not complete.
We have investigated 5 different masks:
\begin{itemize}
\item The vanilla ``Unmasked'' scenario.
\item The one expected from the eROSITA predicted efficiency.
\item The one expected from the PRISM predicted efficiency.
\item A mask that cuts every cluster under $z<0.3$ regardless of its size, used to compare our results with \cite{Creque-Sarbinowski:2016wue}, ``$z>0.3$''.
This is also a futuristic scenario.
\end{itemize}
\label{sec:MaskingProcedure}

\begin{figure}
\centering
\includegraphics[width=0.7\textwidth]{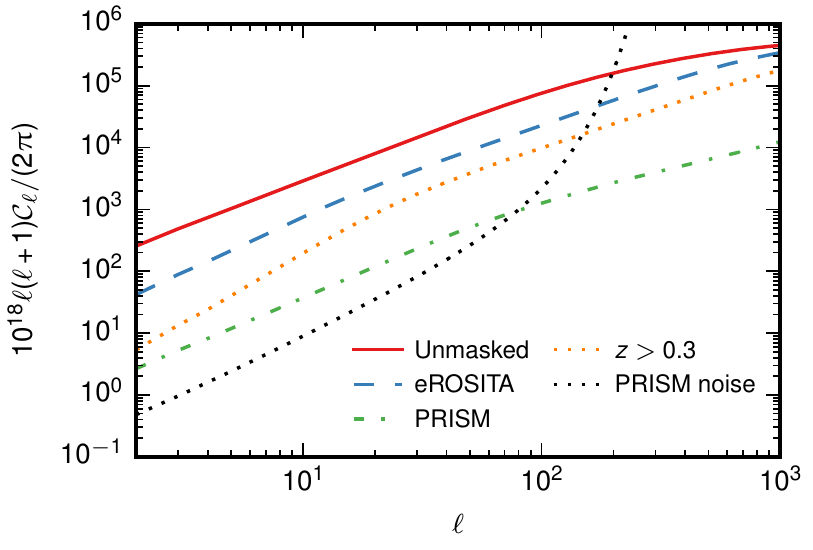}
\caption{SZ power spectrum, for the different sky-masks defined in table \ref{tab:masks}.
For comparison we also plot the PRISM noise level. PIXIE noise level is two orders of magnitude bigger then the PRISM one.
}
\label{fig:MaskComparison}
\end{figure}

The adopted integration boundaries for the various cases are shown in table \ref{tab:masks}.
In figure \ref{fig:MaskComparison} we compare the total (1-halo + 2-halo) SZ power spectra, obtained using different masks.
As shown, using PIXIE with the eROSITA mask, already guarantees a $f_\text{NL}^y$ signal-to-noise ratio comparable with $f_\text{NL}^\mu$. 

\begin{table}
\centering
\begin{tabular}{|cc|}
\hline
Unmasked            & None   \\
eROSITA             & $M<2\times 10^{14}M_\odot/h$ for $z<0.15$    \\
PRISM                & $M<10^{13}M_\odot h^{-1}$\\
$z>0.3$             & $z>0.3$      \\
\hline
\end{tabular}
\caption{Integration boundaries that define the  different masks we use.}
\label{tab:masks}
\end{table}

Masking resolved clusters lowers the noise, but at the same time reduces the level of sky coverage.
We get a rough and conservative estimate of the available fraction of the sky after cluster masking ($f_\text{sky}$) in the following way.
We consider the eROSITA and estimate the number of masked clusters ($<1800$) based on its expected performance \cite{Merloni:2012uf}.
We then assume that the redshift distribution of clusters is constant in redshift for $z\in[0.02,0.15]$; this is a conservative choice since the expected redshift distribution increases rapidly with redshift in the considered range.
Finally we assume that each cluster has a size of 6 Mpc.
This leads to a final estimate
\begin{equation}
f_\text{sky} \approx 1- \frac{1}{0.13} \int_{0.02}^{0.15} \diff z \frac{\pi (6 \text{ Mpc})^2}{4\pi D_A^2(z)}\times 1800  \approx 0.7\ ,
\end{equation}
which will be included in our forecasts.

\subsection{Adding polarization}\label{sec:forecasty}

We consider $y$-E contributions to the signal, by adapting equation (\ref{eq:Tmarg}) into:
\begin{equation}
\textbf{Cov}_{\ell}
=
\frac{1}{2\ell+1}\begin{pmatrix}
C_\ell^{EE}                                                                                 & f^y_\text{NL} C_\ell^{y E} + \alpha_E C_\ell^{\text{SZ} E} +\beta_E C_\ell^{y_\text{reio} E}\\
f^y_\text{NL} C_\ell^{y E} + \alpha_E C_\ell^{\text{SZ} E } +\beta_E C_\ell^{y_\text{reio} E}  & C_\ell^{yy,N} + C_\ell^{1h} + C_\ell^{2h}\\   
\end{pmatrix}.
\end{equation}
The results we get after marginalizing over both $\alpha_E$ and $\beta_E$ are shown in table \ref{tab:results}.
For a PIXIE-like experiment, just replacing $T$ with $E$ tightens the constraints by more than a factor 2.
This comes from two effects. First, the signal-to-noise ratio for $y$-$E$ is intrinsically higher --- just like, and for the same reason as $\mu$-$E$. Second,  marginalizing over secondary signals do not degrades the constraint as much because the primordial and the secondary signal are less correlated as shown in eq. (\ref{eq:SZT-yTcorr}).

We now want to perform a joint analysis of $y$-$T$ and $y$-$E$ signals. 
In this case the covariance is
\begin{equation}
\begin{split}
&(2\ell+1) \textbf{Cov}_{\ell}
=
\\
&=
\begin{pmatrix}
C_\ell^{TT}    & C_\ell^{TE}                                                                                 & f^y_\text{NL} C_\ell^{y T} + \alpha_T C_\ell^{\text{SZ} T}                                                       \\
 C_\ell^{TE}   & C_\ell^{EE}                                                                                 & f^y_\text{NL} C_\ell^{y E} + \alpha_E C_\ell^{\text{SZ} E} +\beta_E C_\ell^{y_\text{reio} E}\\
f^y_\text{NL} C_\ell^{y T} + \alpha_T C_\ell^{\text{SZ} T}& f^y_\text{NL} C_\ell^{y E} + \alpha_E C_\ell^{\text{SZ} E } +\beta_E C_\ell^{y_\text{reio} E}  & C_\ell^{yy,N} + C_\ell^{1h} + C_\ell^{2h}\\   
\end{pmatrix}.
\end{split}
\label{eq:YJointForecast}
\end{equation}

The $C_\ell^{y_\text{reio}E}$ contribution is very small and it has a very different slope than the primordial signal, hence marginalizing over $\beta_E$ changes the final signal-to-noise ratio to a percent level.
For this reason in the joint forecast we fixed $\beta_E=1$.
Moreover theoretical uncertainties in the $SZ$-$T$ and $SZ$-$E$ correlations, which have been parametrized above in terms of the amplitudes $\alpha_T$ and $\alpha_E$, are entirely driven by errors in the prediction of the $y$ signal.
Therefore, we can also assume $\alpha \equiv \alpha_T = \alpha_E$ and marginalize over $\alpha$.

Our joint-analysis results are summarized in table \ref{tab:results} for PIXIE, in \ref{tab:resultsPRISM} for PRISM and in \ref{tab:resultsCV} for an ideal cosmic-variance limited experiment. There we show how adding polarization tightens the constraints both in the PIXIE and PRISM-like, and in the cosmic-variance limited scenario by a factor $\sim 3$

\medskip

\subsection{Cross-correlation with external tracers}
\label{sec:cleaning}

The dominant term in the variance, in all configurations and for both $E$ and $T$, is the SZ power spectrum.
For this reason the only way to further enhance the signal-to-noise ratio, at this stage, is to remove as much contamination from SZ as possible.
Using masks help significantly but, of course, cannot remove the significant contribution from the background of unresolved clusters. 

In order to lower the noise contribution coming from this unresolved background, we consider here an approach based on statistical 
reconstruction of the SZ $y$-map, via correlations with CMB and galaxy-lensing signals. To do this we adapt the method studied in \cite{Manzotti:2014kta}.

In other words, the observed $y$-distortion is the sum of the primordial component, the SZ component and the noise: $y_\text{obs}(\vec{\hat{n}})= y_p(\vec{\hat{n}})+y_\text{SZ}(\vec{\hat{n}})+y_N(\vec{\hat{n}})$. Given the estimate $\hat{y}_\text{SZ}$ of $y_\text{SZ}$, one can use $y_\text{clean}(\vec{\hat{n}}) \equiv y_\text{obs}(\vec{\hat{n}}) -\hat{y}_\text{SZ}(\vec{\hat{n}})$ in place of $y_\text{obs}(\vec{\hat{n}})$.

We will start considering the CMB-lensing-potential, $\phi$, as our SZ-tracer. Later we will reapply the same procedure using the galaxy-lensing convergence field.
The joint probability density function of $y_\text{SZ}$, $\phi$, $T,$ and $E$ is
\begin{equation}
p(\vec{d}_\ell) =  \mathcal{N}( 0, \mathbf{A}_\ell) ,
\end{equation}
where $\mathcal{N}(\vec{\mu}, \mathbf{Cov})$ is the multivariate Normal distribution with mean $\vec{\mu}$ and covariance $ \mathbf{Cov}$, and we defined
\begin{equation}
\mathbf{A}_\ell=
\begin{pmatrix}
C_{\ell}^\text{SZSZ}       & C_{\ell}^{\text{SZ}\phi} & C_{\ell}^{\text{SZ}T} & C_{\ell}^{\text{SZ}E}\\
C_{\ell}^{\text{SZ}\phi} & C_{\ell}^{\phi\phi} &  C_{\ell}^{\phi T}     & C_{\ell}^{\phi E}\\
C_{\ell}^{\text{SZ}T}    & C_{\ell}^{\phi T}   & C_{\ell}^{TT}     & C_{\ell}^{TE}\\
C_{\ell}^{\text{SZ}E}    & C_{\ell}^{\phi E}   & C_{\ell}^{TE}     & C_{\ell}^{EE}
\end{pmatrix}
\equiv
\begin{pmatrix}
C_{\ell}^\text{SZSZ}       & \mathbf{C}_\ell\\
\mathbf{C}^{T}_\ell & \mathbf{B}_\ell
\end{pmatrix},
\qquad
\vec{d}_\ell \equiv
\begin{pmatrix}
a_{\ell m}^\text{SZ}\\
a_{\ell m}^\phi\\
a_{\ell m}^T\\
a_{\ell m}^E
\end{pmatrix}
\equiv
\begin{pmatrix}
a_{\ell m}^\text{SZ}\\
\vec{t}_{\ell m}
\end{pmatrix}.
\label{eq:DodJointLikelihood}
\end{equation}
The conditional probability distribution of the $a_{\ell m}^\text{SZ}$ given the measurement of $\phi$, $T$, and $E$ is
\begin{equation}
p(a_{\ell m}^\text{SZ}|\vec{t}_{\ell m}) = \mathcal{N}(\mathbf{C}^T_\ell \mathbf{B}^{-1}_\ell \vec{t}_{\ell m}, C_\ell^\text{SZSZ}- \mathbf{C}_\ell^T \mathbf{B}^{-1}_\ell  \mathbf{C}_\ell ).
\end{equation}
The expectation value of $a_{\ell m}^\text{SZ}$ then is
\begin{equation}
\hat{a}_{\ell m}^\text{SZ} = \mathbf{C}^T_\ell \mathbf{B}^{-1}_\ell \vec{t}_{\ell m}.
\label{eq:haty}
\end{equation}
The probability distribution of $ a_{\ell m}^\text{clean}= a_{\ell m}^\text{obs}  - \hat{a}_{\ell m}^\text{SZ} $ is
\begin{equation}
p (a_{\ell m}^\text{clean}|\vec{t}_{\ell m}) = \mathcal{N} \left(-\mathbf{C}^T_\ell \mathbf{B}^{-1}_\ell \vec{t}_{\ell m}, Var(a_{\ell m}^\text{obs}) + Var(\hat{a}_{\ell m}^\text{SZ}) -2 Cov(a_{\ell m}^\text{obs}  , \hat{a}_{\ell m}^\text{SZ}) \right)   .
\label{eq:acleancondprob}
\end{equation}
The covariance of $a_{\ell m}^\text{obs}$ and $\hat{a}_{\ell m}^\text{SZ}$ can be computed using eq. (\ref{eq:haty}) to write $a_{\ell m}^\text{SZ}$ in term of the multipolar coefficients of the tracers.
For our fiducial model $f_\text{NL}=0$ the only term in $a_{\ell m}^\text{obs}$ that does contribute to the cross correlation is indeed $a_{\ell m}^\text{SZ}$.

Note that the non-zero mean appearing in eq. (\ref{eq:acleancondprob}) is due to the fact that we are considering the probability of $a_{\ell m}^\text{clean}$, conditional to the specific observed realization of $\vec{t}_{\ell m}$ in the sky. Of course, if one consider the expectation value over the ensemble of possible realization, one recovers $\langle \vec{t}_{\ell m} \rangle = 0$ and therefore $\langle a_{\ell m}^\text{clean} \rangle = 0$ as expected.

Note also that, neglecting temperature and polarization, eq. (\ref{eq:haty}) simply becomes $\hat{a}_{\ell m}^\text{SZ} = (C_\ell^{\phi \text{SZ}}/C_\ell^{\phi \phi}) a_{\ell m}^\phi$, as expected.

Using $y_\text{clean}(\vec{\hat{n}})$ instead of $y_\text{obs}(\vec{\hat{n}})$ leaves the numerator of eq. (\ref{eq:YJointForecast}) unchanged. The variance instead is modified, with the following replacement, which of course lowers the overall SZ-noise contribution:
\begin{equation}
(C_\ell^{1h} + C_\ell^{2h})\rightarrow\langle (a_{\ell m}^\text{SZ}-\hat{a}_{\ell m})^2\rangle = (C_\ell^{1h} + C_\ell^{2h}) - \langle \hat{a}^\text{SZ}_{\ell m} \hat{a}_{\ell m'}^\text{SZ}\rangle.
\label{eq:NoiseReduction}
\end{equation}
The structure of the 1-halo and 2-halo terms for the CMB lensing potential cross-correlation with the SZ effect is the same as for the SZ power spectrum, reading \cite{Hill:2013dxa}:
\begin{gather}
\nonumber
C_\ell^{\text{SZ}\phi,1h}  = \int \diff z \frac{\Diff{2}V}{\diff x \diff \Omega} \int \diff M \frac{\diff n}{\diff M} (z, M) | \tilde{y}_\ell(z,M) \tilde{\phi}_\ell(z,M)|
\\
\begin{split}
C_\ell^{\text{SZ}\phi,2h}  = \int \diff z \frac{\Diff{2}V}{\diff x \diff \Omega} D_+^2(z) P_m(k)
&
\bigg[\int \diff M \frac{\diff n}{\diff M} (z, M)  b(z, M) a(z) \tilde{y}_{3D}(z,M,k) \bigg]\times
\\
\times
&
\bigg[\int \diff M \frac{\diff n}{\diff M} (z, M)  b(z, M) a(z) \tilde{\phi}_{3D}(z,M,k) \bigg
]\bigg|_{k=\big(\frac{\ell + 1/2}{\chi(z)}\big)},
\end{split}
\label{eq:phiSZ}
\end{gather}
where $\tilde{\phi}_{3D}(z,M,k)$ is the Fourier transform of the halo contribution to the projected lensing potential,
\begin{equation}
\tilde{\phi}_\ell(z,M) =\frac{2}{\ell(\ell+1)} \frac{4\pi r_{s,\phi}}{\ell_{s,\phi}^2} \int \diff x x^2 j_0\left(\frac{k x}{\ell_s}\right)\frac{4 \pi G \chi(z) (\chi_* -\chi(z))\rho_\text{NFW}(z,M,k)}{c^2 \chi_* (1+z)},
\end{equation}
$r_{s,\phi}$ is the typical scale radius for $\tilde{\phi}_{3D}(z,M,k)$  and $\ell_{s,\phi} = a(z) \chi(z)/r_{s,\phi} $ is the associated multipole.
One gets the CMB lensing potential power spectrum replacing the remaining $\tilde{y}_{\ell}(z,M)$ with $\tilde{\phi}_{\ell}(z,M)$.
We checked our spectra against those shown in \cite{Hill:2013dxa},
changing our integration boundaries to match their choices, and we are in very good agreement with them.

As discussed in \cite{Hill:2013dxa} the correlation of SZ and $\phi$ is small ($\approx 0.3-0.4$) up to $\ell = \text{few}\times10^3$ so we might expect only a small improvement.
As a zeroth-order approximation, we can neglect $T$ and $E$ in eq. (\ref{eq:haty}) and write $\hat{a}^\text{SZ}_{\ell m} \approx 0.4 \times a^\text{SZ}_{\ell m}$.
In this limit the relation in eq. (\ref{eq:NoiseReduction}) becomes
\begin{equation}
(C_\ell^{1h} + C_\ell^{2h})\rightarrow (C_\ell^{1h} + C_\ell^{2h}) - \langle \hat{a}^\text{SZ}_{\ell m} \hat{a}_{\ell m'}^\text{SZ}\rangle \approx 0.84 (C_\ell^{1h} + C_\ell^{2h}).
\end{equation}
Therefore using CMB lensing as a tracer should provide a $\approx10\%$ improvement.
Indeed the numerical evaluation of the cross-correlations validates this back-of-the-envelope estimate.
\\

The second tracer we investigate is the galaxy-lensing convergence field.
Its cross correlation with the SZ effect can again be computed by replacing, in eq. (\ref{eq:phiSZ}), $\tilde{\phi}_\ell(z,M)$ with \cite{VanWaerbeke:2013cfa, Ma:2014dea}
\begin{equation}
\tilde{\kappa}_\ell(z,M) = \frac{4\pi r_{s,\kappa}}{\ell_{s,\kappa}^2} \int \diff x x^2 j_0\left(\frac{k x}{\ell_{s, \kappa}}\right)\frac{4 \pi G g(z) \rho_\text{NFW}(z,M,k)}{c^2 (1+z)},
\end{equation}
where $r_{s,\kappa}$ is the typical scale radius of the lensing potential of the halo and $\ell_s = a(z) \chi(z)/r_{s,\kappa} $, and we defined
\begin{equation}
g(z) \equiv \int_{\chi (z)}^\infty \diff \chi' \frac{\chi(z) [\chi' -\chi(z)]}{\chi'} p_S(\chi'),
\end{equation}
where $p_S$ is the redshift distribution of the sources.
Again, one gets the power spectrum replacing the remaining $\tilde{y}_{\ell}(z,M)$ with $\tilde{\kappa}_{\ell}(z,M)$.

In this case, due to the higher correlation between the SZ effect and the galaxy lensing, the cleaning procedure performs better than with the CMB lensing.
However the signal-to-noise ratio achieved with this procedure alone is still smaller than the S/N achievable via direct cluster masking.
The optimal way to proceed is therefore to adopt the two approaches in combination.
This can be done by reconstructing the $\hat{y}_\text{SZ}(\vec{\hat{n}})$ map using tracers as discussed; 
then the resolved clusters can be masked in both the $y_\text{obs}(\vec{\hat{n}})$ and the $\hat{y}_\text{SZ}(\vec{\hat{n}})$ maps.
The $y_\text{clean}(\vec{\hat{n}})$ masked map is then obtained by difference.

We model this procedure in our forecast by changing the integration boundaries of all the integrals involving at least one power of $\tilde{y}_\ell(z,M)$ as discussed in section \ref{sec:MaskingProcedure}.
The integrals involving only powers of $\tilde{\kappa}_\ell(z,M)$ (e.g. the second square bracket in the 2-halo term, eq. (\ref{eq:phiSZ})) are left unmodified as the mask is applied to the reconstructed $\hat{y}_\text{SZ}(\vec{\hat{n}})$ map, and not to the input lensing map. The final results are shown in table \ref{tab:results}, for PIXIE, in table \ref{tab:resultsPRISM} for PRISM and in table \ref{tab:resultsCV}, for an ideal survey.

Considering PIXIE, the forecasted signal-to-noise ratio quickly saturate when using more and more futuristic masks in combination with galaxy lensing, because the SZ power spectrum becomes rapidly sub-dominant with respect to the PIXIE noise. In the cosmic-variance limited case, however, SZ remains by far the dominant source of noise, even after cleaning, and makes $y$-based constraints much worse than $\mu$-based one. Nonetheless it is important to stress again that $y$ and $\mu$ probe very different scales.

The measurement of $\fnl^y$ that PRISM will achieve, contrary to PIXIE, won't be significantly limited by instrumental noise.
In fact the signal-to-noise ratio for PRISM is effectively the same of a cosmic-variance limited experiment.
This constraint ($1\sigma(\fnl^y) = 260$ for PRISM) might not appear significant compared with the current bound set by \textit{Planck} ($\fnl = 0.6 \pm 5.0$, 68\% C.L.) at first glance.
However if one consider that $\fnl$ might have a running, its importance change considerably.
For example, if we consider a primordial bispectrum  of the form \cite{Byrnes:2010ft}
\begin{equation}
B(k_1,k_2,k_3)\propto \fnl^*\left[P(k_1)P(k2)\left(\frac{k_3}{k^*}\right)^{n_\text{NG}} + 2 \text{ perm.}\right],
\label{eq:FredBispectrum}
\end{equation}
and use $1\sigma$ upper bounds $\fnl^*=5$, $n_\text{NG}=1$, consistent with current observations \cite{Becker:2012je}, we would expect $\fnl^y \approx 700$ on the $y$-scales, way above the detectability limit.
Even though the bispectrum in eq. (\ref{eq:FredBispectrum}) is theoretically well-motivated, it has to be considered here just as a toy-model, because we made a choice of values of the parameters that might be outside the range of validity of the model itself.
The point here is just to use a phenomenological, toy-model shape, just to show in a simple, quantitative way how $y$-constraints are useful, even if they turn out $2$ orders of magnitude worse than current $T$, $E$ bispectrum bounds.

Of course one may argue that the same holds, even more so, for $\fnl^\mu$ on the $\mu$-scales, but to avoid pathologically large non-Gaussianity on the smallest scales, the increasing trend has to stop somewhere.
Therefore it is again important to study both the $y$- and the $\mu$-scales.

\section{Conclusions}\label{sec:Conclusions}
In this paper we investigated in detail the effects of including CMB polarization in NG studies of cross-correlation between CMB primary anisotropies and $\mu$- and $y$-CMB-distortions. 
Including the previously unaccounted $y$-$E$ spectrum, besides adding new signal, has the important advantage of making the primordial NG analysis more robust, since it removes the large bias arising from the ISW-SZ contribution in the $y$-$T$ spectrum.
Potential spurious contamination in the primordial $y$-$E$ signal can come from reionization, but this turned out to be negligible after a complete numerical analysis at second order in the perturbations.
In addition to considering $y$-$E$ spectra, we also studied in detail how to reduce SZ contamination, thus lowering the overall noise contribution, considering two approaches. The former, already considered in previous works \cite{Creque-Sarbinowski:2016wue, Hill:2013baa}, consists in masking low-redshift 
clusters, detected via X-ray surveys. To this, we add  the exploitation of cross-correlation with external tracers, namely CMB and galaxy lensing, as a way to partially reconstruct the $y$ contribution from unresolved clusters. 
The template so-obtained is then used to clean the $y$-map from the remaining unresolved contribution.

Using this procedure, we obtain $\fnl^y$ forecasts for PIXIE, PRISM, and for an ideal cosmic-variance limited experiment. In all cases we find that including $y$-$E$ leads to improvements in $\fnl^y$ constraints up to a factor $\sim 2$
assuming to mask resolved clusters. A further error bar improvement of order $25 \%$ is expected from external-tracer cross-correlation and template cleaning.
Our final forecasts are then $1\sigma(\fnl^y) = 2300$ for PIXIE, $1\sigma(\fnl^y) = 300$ for PRISM and $1\sigma(\fnl^y) = 130$ for the cosmic-variance limited case.

It is clear that, even in the ideal scenario, $\fnl$ constraints based on $y$ are very poor when compared to current  {\em Planck} bispectrum measurements.
For the cosmic-variance limited case, the errors achievable using $\mu$ are also orders of magnitude smaller than those achievable using $y$.
This is due to residual SZ contamination, still significant even after masking and template reconstruction and cleaning.
Nevertheless, two things are worth noticing:
first, the constrains on $\fnl^y$ and $\fnl^\mu$ achievable with a realistic (not cosmic variance-limited) survey design (e.g. PIXIE, PRISM) are comparable.

Second, and most important, the main goal we consider here is to test NG scale-dependence, In this respect, $\fnl^y$ measurements are very interesting, even with all the limitations imposed by SZ contamination, because they open a new window on an otherwise inaccessible range of scales: a simple example to illustrate this point is provided by the bispectrum toy model, characterized by an $\fnl$-running parameter, considered at the end of section \ref{sec:cleaning}.

\acknowledgments
The authors thank Simeon Bird and Marc Kamionkowski for useful discussions, and Christian Fidler for help with the second order Boltzmann integrator \texttt{SONG}, which was employed in this work to compute $y$-E contributions from reionization.
The authors also thank Jens Chluba for valuable comments and feedback on the draft of the paper.
MS is supported in part by a Grant-in-Aid for JSPS Research under Grant No. 27-10917, and in part by the World Premier International Research
Center Initiative (WPI Initiative), MEXT, Japan.
MS also acknowledges Center for Computational Astrophysics, National Astronomical Observatory of Japan, for providing computing resources of Cray XC30.
NB \& ML acknowledge financial support by ASI Grant 2016-24-H.0.
NB \& ML acknowledge partial financial support by the ASI/INAF Agreement I/072/09/0 for the Planck LFI Activity of Phase E2. 
CloudVeneto is acknowledged for the use of computing and storage facilities.

\appendix
\section{$y_\text{reio}^{(2)}\text{-}E^{(2)}$ cross-correlation}
\label{sec:yE2fullcalculation}

Let's consider a perturbation field $X$ that can be written as an expansion over the primordial density perturbation field $\zeta$. Up to second order, in term of its linear $\mathcal{T}_{X\ \ell m}^{(1)}$ and second order $\mathcal{T}_{X\ \ell m}^{(2)}$ transfer functions the field projection on the sphere can be written as \cite{pettinari:2015a}
\begin{equation}
a_{\ell m}^X(\vec{k})=
\mathcal{T}_{X\ \ell m}^{(1)}(\vec{k}) \zeta(\vec{k})+
\int\frac{\Diff{3} \vec{q_1}\Diff{3}\vec{q_2}}{(2\pi)^3} \delta^{(3)}(\vec{k}-\vec{q_1}-\vec{q_2})
\mathcal{T}_{X\ \ell m}^{(2)}(\vec{q_1},\vec{q_2}, \vec{k})
\zeta(\vec{q_1})\zeta(\vec{q_2}) + ...
\end{equation}
In full generality, the cross-correlation of the second order contributions of two field $X$ and $Y$ will be
\begin{equation}
\begin{split}
\Braket{a_{\ell m}^{X(2)} a_{\ell' m'}^{Y(2)}}=
\int \frac{\Diff{3} \vec{k_1}}{(2\pi)^3} \int & \frac{\Diff{3} \vec{q_1}\Diff{3}\vec{q_2}}{(2\pi)^3}
\mathcal{T}_{X\ \ell m}^{(2)}(\vec{q_1},\vec{q_2}, \vec{k_1})
\delta^{(3)}(\vec{k_1}-\vec{q_1}-\vec{q_2})
\\
\int \frac{\Diff{3} \vec{k_2}}{(2\pi)^3} \int & \frac{\Diff{3} \vec{p_1}\Diff{3}\vec{p_2}}{(2\pi)^3}
\mathcal{T}_{Y\ \ell' m'}^{(2)}(\vec{p_1}, \vec{p_2}, \vec{k_2})
\delta^{(3)}(\vec{k_2}-\vec{p_1}-\vec{p_2})
\\
&\Braket{\zeta(\vec{q_1})\zeta(\vec{q_2})\zeta(\vec{p_1})
\zeta(\vec{p_2})}.
\end{split}
\end{equation}

Using Wick theorem, under the assumption that the primordial perturbation field is Gaussian, and using the fact that this expression in symmetric in $\vec{q_1} \leftrightarrow\vec{q_2}$ one gets
\begin{equation}
\Braket{\zeta(\vec{q_1})\zeta(\vec{q_2})\zeta(\vec{p_1})
\zeta(\vec{p_2})}=2(2\pi)^6 \delta^{(3)}(\vec{q_1}+\vec{p_1}) \delta^{(3)}(\vec{q_2}+\vec{p_2})P(q_1)P(q_2).
\end{equation}

\begin{equation}
\begin{split}
\Braket{a_{\ell m}^{X} a_{\ell' m'}^Y}=2\int & \frac{\Diff{3} \vec{k_1}}{(2\pi)^3} \int \frac{\Diff{3} \vec{q_1}\Diff{3}\vec{q_2}}{(2\pi)^3}
\mathcal{T}_{X\ \ell m}^{(2)}(\vec{q_1},\vec{q_2}, \vec{k_1})
\mathcal{T}_{Y\ \ell' m'}^{(2)}(-\vec{q_1}, -\vec{q_2}, \vec{-k_1})
\\
&
\delta^{(3)}(\vec{k_1}-\vec{q_1}-\vec{q_2})P(q_1)P(q_2).
\end{split}
\label{eq.RawCorr}
\end{equation}

If we assume rotational invariance we can rotate our reference system to match $\hat{\vec{z}}$ with the direction of $\vec{k}$.
\begin{equation}
\begin{split}
\Braket{a_{\ell m}^{X} a_{\ell' m'}^Y}=
&
2
\int \frac{\Diff{3} \vec{q_1}\Diff{3}\vec{q_2}}{(2\pi)^3}
\int \frac{k_1^2\diff k_1}{(2\pi)^3}
\sqrt{\frac{4\pi}{(2\ell+1)}}
\sqrt{\frac{4\pi}{(2\ell'+1)}}
\\
&
\sum_{m_1m_2}
\mathcal{T}_{X\ \ell m_1}^{(2)}(\vec{q_1},\vec{q_2}, k_1)
\mathcal{T}_{Y\ \ell' m_2}^{(2)}(-\vec{q_1},- \vec{q_2}, k_1)
\\
&\delta^{(3)}(-\vec{q_1}-\vec{q_2}+k_1\vec{\hat{z}})
P(q_1)P(q_2)
\int \Diff{2} \Omega(\vec{\hat{k}_1})(-1)^{\ell'}
\  _{-m_1}Y_\ell^{m} (\vec{\hat{k}_1})\  _{m_2}Y_{\ell'}^{m'} (\vec{\hat{k}_1})=
\\
=&
\frac{8\pi}{2\ell+1}
\int \frac{\Diff{3} \vec{q_1}\Diff{3}\vec{q_2}}{(2\pi)^6}
\int k_1^2\diff k_1
\delta^{(3)}(-\vec{q_1}-\vec{q_2}+k_1\vec{\hat{z}})
P(q_1)P(q_2)
\\
&
\sum_{m_1}
\mathcal{T}_{X\ \ell m_1}^{(2)}(\vec{q_1},\vec{q_2}, k_1)
\mathcal{T}_{Y\ \ell' m_1}^{(2)}(-\vec{q_1}, -\vec{q_2}, k_1)
(-1)^{l'+m_1-m}\delta_{\ell}^{\ell'}\delta_{m_1}^{m_2}\delta_{m}^{-m'}.
\end{split}
\end{equation}
In the first line we used the fact that the transfer functions transform under rotations as spherical harmonics; and in the second $\int \Diff{2} \Omega(\vec{\hat{k}_1})\  _{-m_1}Y_\ell^{m} (\vec{\hat{k}_1})\  _{-m_2}Y_{\ell'}^{m'} (\vec{\hat{k}_1})= (-1)^{m1+m} \delta_{\ell}^{\ell'}\delta_{m_1}^{-m_2}\delta_{m}^{-m'}$.

Now we specialize in the case we are interested in: $X$ being the CMB polarization and $Y$ being the quadratic Doppler effect effect. To uniform our notation with \cite{pettinari:2015a} and factor out the quantities that \texttt{SONG} actually calculates $\overline{\mathcal{T}}$, we perform the substitution
\begin{equation}
\mathcal{T}_{X\ \ell m_1}^{(2)}(\vec{q_1},\vec{q_2}, k_1)
=
(-1)^{m_1}\sqrt{\frac{4\pi}{2|m_1| +1}}
\overline{\mathcal{T}}_{X\ \ell m_1}^{(2)}(q_1, q_2, k_1)
Y_{|m_1|}^{m_1}(\vec{\hat{q}_1})
\end{equation}
\begin{equation}
\begin{split}
\Braket{a_{\ell m}^{X} a_{\ell' m'}^Y}=&
\delta_{\ell}^{\ell'}\delta_{m}^{-m'}
\frac{8\pi}{2\ell+1}
\int \frac{\Diff{3} \vec{q_1}\Diff{3}\vec{q_2}}{(2\pi)^6}
\int k_1^2\diff k_1
\delta^{(3)}(\vec{k_1}-\vec{q_1}-\vec{q_2})
P(q_1)P(q_2)
\\
&
\sum_{m_1}
\sqrt{\frac{4\pi}{2|m_1| +1}}(-1)^{\ell'+2m_1-m}
\overline{\mathcal{T}}_{X\ \ell m_1}^{(2)}(q_1, q_2, k_1)
Y_{|m_1|}^{m_1}(\vec{\hat{q}_1})
\mathcal{T}_{Y\ \ell' m_1}^{(2)}(-\vec{q_1}, -\vec{q_2}, k_1)
.
\end{split}
\end{equation}
For spectral distortion from reionization \cite{Renaux-Petel:2013zwa}
\begin{equation}
\begin{split}
\mathcal{T}_{Y\ \ell' m_1}^{(2)}(\vec{q_1}, \vec{q_2}, k_1) = &
(2\ell' + 1)  \bigg[\frac{-\delta^0_{m_1}}{3} I_{\ell'}^{(1)}(q_1,q_2, k_1)\ \vec{\hat{q}_1}\cdot\vec{\hat{q}_2} +
\\
&
+\frac{11\pi}{45}I_{\ell',m_1}^{(2)}(q_1,q_2, k_1)\sum^1_{n=-1} \alpha_{n,m_1}\big(Y^{-m_1-n}_1(\vec{\hat{q}_1})Y^{n}_1(\vec{\hat{q}_2})\big)^{\!*}\bigg]
\end{split}
\end{equation}
with
\[
\begin{split}
I_{\ell'}^{(1)}(q_1,q_2, k_1)=&
\int^{\eta_0}_{\eta_{reio}} \diff \eta \ g(\eta)j_\ell(k_1 r(\eta))F(q_1,\eta)F(q_2,\eta)
\\
I_{\ell',m_1}^{(2)}(q_1,q_2, k_1)=&
\int^{\eta_0}_{\eta_{reio}} \diff \eta \ g(\eta)
j_\ell^{(2, m_1)}(k_1 r(\eta))F(q_1,\eta)F(q_2,\eta).
\end{split}
\]
\[
\vec{\hat{q}_1}\cdot\vec{\hat{q}_2}= \frac{4\pi}{3}\sum_{m_2=-1}^{1} Y_1^{m_2*}(\vec{\hat{q}_1})Y_1^{m_2}(\vec{\hat{q}_2})
\]
and
\[
\alpha_{0,m}\equiv \sqrt{4-m^2} \quad \alpha_{\pm 1,m}\equiv \sqrt{(2\pm m)(2\pm m-1)/2}.
\]
Here $F(k, \eta)$ is the baryon velocity transfer function, defined as 
\begin{equation}
\vec{v}_b (k, \eta) = -i \frac{\vec{k}}{k} F(k,\eta) \zeta(\vec{k})\; ,
\end{equation}
and $g(\eta)$ is the visibility function.

Plugging everything back in, and using the Reylight expansion of the Dirac delta we get
\begin{equation}
\begin{split}
\Braket{a_{\ell m}^{X} a_{\ell' m'}^Y}
=
&
\delta_{\ell}^{\ell'}\delta_{m}^{-m'}
\frac{8\pi}{2\ell+1}
\int \frac{\Diff{3} \vec{q_1}\Diff{3}\vec{q_2}}{(2\pi)^6}
\int k_1^2\diff k_1
8
\int x^2 \diff x 
\sum_{L}\sum_{L_1M_1}\sum_{L_2M_2}
j_{L}(x k_1)j_{L_1}(x q_1)j_{L_2}(x q_2)
\\
&
\sqrt{\frac{2L+1}{4\pi}}
(-1)^{L_1+L_2}
Y_{L_1}^{M_1*}(\vec{\hat{q}_1})
Y_{L_2}^{M_2*}(\vec{\hat{q}_2})
i^{L+L_1+L_2}
h_{L_1L_2L}
\threej{L_1}{L_2}{L}{M_1}{M_2}{0}
\\
&
\sum_{m_1}
\sqrt{\frac{4\pi}{2|m_1| +1}}(-1)^{\ell'+2m_1-m}
\overline{\mathcal{T}}_{X\ \ell m_1}^{(2)}(q_1, q_2, k_1)
Y_{|m_1|}^{m_1}(\vec{\hat{q}_1})
\\
&
(2\ell' + 1)  \bigg[\frac{-\delta^0_{m_1}}{3} I_{\ell'}^{(1)}(q_1,q_2, k_1)\ \frac{4\pi}{3}\sum_{m_2=-1}^{1} Y_1^{m_2*}(\vec{\hat{q}_1})Y_1^{m_2}(\vec{\hat{q}_2}) +
\\
&
+\frac{11\pi}{45}I_{\ell',m_1}^{(2)}(q_1,q_2, k_1)\sum^1_{n=-1} \alpha_{n,m_1}\big(Y^{m_1-n}_1(\vec{\hat{q}_1})Y^{n}_1(\vec{\hat{q}_2})\big)^{\!*}\bigg]
P(q_1)P(q_2)=
\\
=
&
\delta_{\ell}^{\ell'}\delta_{m}^{-m'}
\frac{8\pi}{2\ell+1}
\int \frac{q_1^2 \diff q_1 q_2^2 \diff q_2}{(2\pi)^6}
\int k_1^2\diff k_1
8
\int x^2 \diff x 
\sum_{L}\sum_{L_1M_1}\sum_{L_2M_2}
j_{L}(x k_1)j_{L_1}(x q_1)j_{L_2}(x q_2)
\\
&
\sqrt{\frac{2L+1}{4\pi}}
(-1)^{L_1+L_2}
i^{L+L_1+L_2}
h_{L_1L_2L}
\threej{L_1}{L_2}{L}{M_1}{M_2}{0}
\sum_{m_1}
\sqrt{\frac{4\pi}{2|m_1| +1}}(-1)^{\ell'+2m_1-m}
\\
&
\overline{\mathcal{T}}_{X\ \ell m_1}^{(2)}(q_1, q_2, k_1)
(2\ell' + 1)  \bigg[\frac{-\delta^0_{m_1}}{3} I_{\ell'}^{(1)}(q_1,q_2, k_1)\ \frac{4\pi}{3}\sum_{m_2=-1}^{1} (-1)^{m_2} \delta_{L_2}^1 \delta_{M_2}^{m_2} \frac{3}{\sqrt{4\pi}}\delta^1_{L_1}\delta_{M_1}^{-m_2}+
\\
&
\frac{11\pi}{45}I_{\ell',m_1}^{(2)}(q_1,q_2, k_1)\sum^1_{n=-1} \alpha_{n,m_1}
h_{L_1|m_1|1}
\threej{L_1}{|m_1|}{1}{-M1}{m_1}{-m_1+n}
\\
&
(-1)^{M_1+m_1-n}\delta_{L_2}^1\delta_{M_2}^{-n}
\bigg]
P(q_1)P(q_2)
\end{split}
\end{equation}
and as usual
\[
h_{L_1L_2L_3} = \sqrt{\frac{(2L+1)(2L_1+1)(2L_2+1)}{4\pi}}\threej{L}{L_1}{L_2}{0}{0}{0}.
\]

The first term is further reduced to
\begin{equation}
\begin{split}
1st=
&(-1)^{\ell'-m}\,
\delta_{\ell}^{\ell'}\delta_{m}^{-m'}
64\pi\!
\int \frac{q_1^2 \diff q_1 q_2^2 \diff q_2}{(2\pi)^6}
\int k_1^2\diff k_1
\int x^2 \diff x 
j_{0}(x k_1)j_{1}(x q_1)j_{1}(x q_2)
\\
&
\overline{\mathcal{T}}_{X\ \ell, 0}^{(2)}(q_1, q_2, k_1)
\frac{1}{3}I_{\ell'}^{(1)}(q_1,q_2, k_1)
P(q_1)P(q_2)
\\
\end{split}
\end{equation}
where we used the relation $\sum_{m_2=-1}^{1}(-1)^{m_2}\threej{1}{1}{L}{-m_2}{m_2}{0}=-\sqrt{3}\delta_L^0;$
whereas the second is
\begin{equation}
\begin{split}
2nd=
&
(-1)^{\ell'-m}
\delta_{\ell}^{\ell'}\delta_{m}^{-m'}
64\pi\!
\int \frac{q_1^2 \diff q_1 q_2^2 \diff q_2}{(2\pi)^6}
\int k_1^2\diff k_1
\sum_{L}\sum_{L_1}\sum^1_{n=-1} 
\int x^2 \diff x 
j_{L}(x k_1)j_{L_1}(x q_1)j_{1}(x q_2)
\\
&
\sum_{m_1}
\overline{\mathcal{T}}_{X\ \ell m_1}^{(2)}(q_1, q_2, k_1)
\frac{11\pi}{45}I_{\ell',m_1}^{(2)}(q_1,q_2, k_1)
P(q_1)P(q_2)
\\
&
(-1)^{L_1+1}
i^{L+L_1+1}
(-1)^{3m_1}
\frac{3(2L+1)(2L_1+1)}{4\pi}
\alpha_{n,m_1}
\threej{L_1}{1}{L}{0}{0}{0}
\threej{L_1}{1}{L}{n}{-n}{0}
\\
&
\threej{L_1}{|m_1|}{1}{0}{0}{0}
\threej{L_1}{|m_1|}{1}{-n}{m_1}{-m_1+n}
\end{split}
\end{equation}
The structure of the 4 three-$j$ symbol that appear in the second term guarantees that the sum over $L,L_1$ is not infinite. In fact we found that their product is non zero only for $L<4, L_1<3$.

As the integral over $x$, for both terms, has to be computed for small values of the multipolar indices, it can be evaluated analytically using the relation \cite{0305-4470-24-7-018}
\begin{equation}
\begin{split}
&
I(L_1,L_2,L_3,q_1,q_2,k)
\equiv 
\int x^2 \diff x 
j_{L_1}(x q_1)j_{L_2}(x q_2) j_{L}(x k_1)
=
\\
=
&
\frac{\pi \beta(\Delta)}{4 q_1 q_2 k}i^{L_1+L_2-L}\sqrt{2L +1}\bigg(\frac{q_1}{k}\bigg)^L
\threej{L_1}{L_2}{L}{0}{0}{0}^{-1}
\\
&
\sum_{\mathcal{L}=0}^L
\begin{pmatrix}
2L\\
2\mathcal{L}
\end{pmatrix}^{1/2}
\bigg(\frac{q_2}{q_1}\bigg)^\mathcal{L}
\sum_l
(2l+1)
\threej{L_1}{L-\mathcal{L}}{l}{0}{0}{0}
\threej{L_2}{\mathcal{L}}{l}{0}{0}{0}
\sixj{L_1}{L_2}{L}{\mathcal{L}}{L-\mathcal{L}}{l}
P_l(\Delta).
\end{split}
\label{eq:Int3j}
\end{equation}
The triangular condition over the three momenta is enforced by $\beta(x)=\theta_H(1-x)\theta_H(1+x)$, where $\Delta=\frac{q_1^2+q_2^2-k^2}{2q_1 q_2}$ and $\theta_H(x)$ is the modified Heaviside function.

Wrapping up we get
\begin{equation}
\begin{split}
y_\text{reio}^{(2)}\text{-}E^{(2)} = 
&(-1)^{\ell'-m}64\pi
\int \frac{q_1^2 \diff q_1 q_2^2 \diff q_2}{(2\pi)^3}
\int \frac{k_1^2\diff k_1}{(2\pi)^3}
P(q_1)P(q_2)\delta_{\ell}^{\ell'}\delta_{m}^{-m'}
\\
&
\bigg[
\overline{\mathcal{T}}_{X\ \ell, 0}^{(2)}(q_1, q_2, k_1)
\frac{1}{3}I_{\ell'}^{(1)}(q_1,q_2, k_1)
I(0,1,1,k_1,q_1,q_2)
+
\\
&
+\sum_{L}^4
\sum_{L_1}^3
\sum_{m_1}\sum^1_{n=-1} 
(-1)^{L_1+1}
i^{L+L_1+1}
(-1)^{3m_1}
\frac{3(2L+1)(2L_1+1)}{4\pi}
\alpha_{n,m_1}
\\
&
\threej{L_1}{1}{L}{0}{0}{0}
\threej{L_1}{1}{L}{n}{-n}{0}
\threej{L_1}{|m_1|}{1}{0}{0}{0}
\threej{L_1}{|m_1|}{1}{-n}{m_1}{-m_1+n}
\\
&
\frac{11\pi}{45}I_{\ell',m_1}^{(2)}(q_1,q_2, k_1)
\overline{\mathcal{T}}_{X\ \ell m_1}^{(2)}(q_1, q_2, k_1)
I(L,L_1,1,k_1,q_1,q_2)
\bigg].
\end{split}
\end{equation}
While the angular --- moments independent --- part can be computed analytically, the integrals over $q_1$, $q_2$, and $k_1$ have to be evaluated numerically.
Luckily enough the structure of these integrals is the same one finds when calculating the intrinsic bispectrum of the CMB.
Therefore \cite{pettinari:2014a} provides a good insight of what are the properties of the integrand.
In fact we found that it oscillates both along $k_1$, and along $q_1$ and $q_2$, however the frequency of oscillation along $k_1$ is one order of magnitude higher than the other.
For this reason we computed the $k_1$ integral over a coarse grid of $q_1$ and  $q_2$, and only then we performed the integral of the now smoother function.
Moreover the symmetry $\vec{q_1}\leftrightarrow\vec{q_2}$ allows us to pick only the configurations with $q_2 < q_1$ and double the result of the integral in the end.

\bibliographystyle{JHEP}

\bibliography{bibliografia}

\end{document}